\documentclass[%
 reprint,
 amsmath,amssymb,
 aps,
]{revtex4-2}
\DeclareUnicodeCharacter{202F}{\,}
\usepackage{graphicx}% Include figure files
\usepackage{dcolumn}% Align table columns on decimal point
\usepackage{enumitem} 
\usepackage{bm}% bold math

\begin{document}

\preprint{APS/123-QED}

\title{Mechanism of the quasi-elastic scattering based on the dinuclear system concept}

\author{Zehong Liao}
\author{Yu Yang}
\author{Zepeng Gao}
\author{Jun Su}
\author{Long Zhu}
 \email{Contact author: zhulong@mail.sysu.edu.cn}

%\author{Yueping Fang$^{1}$}%
 \affiliation{Sino-French Institute of Nuclear Engineering and Technology, Sun Yat-sen University, Zhuhai 519082, China}

\date{\today}

\begin{abstract}
%\begin{description}
A unified description of full reaction channels in low-energy heavy-ion collisions is a great challenge. Although, the theoretical models based on the dinuclear system (DNS) concept have been successfully employed in multinucleon transfer (MNT) reactions, the underestimation of the quasi-elastic (QE) channel results in unreliable description of few nucleon transfer, especially for the light reaction systems. In this work, the DNS-sysu model is improved by introducing the impact-parameter-dependent transition probabilities for a unified description of few nucleon and many nucleon transfer in MNT reactions. Extensive experimental data—including reactions such as $^{40}$Ca, $^{58, 64}$Ni, $^{136}$Xe, and $^{208}$Pb + $^{208}$Pb were compared with the model predictions. The calculated isotopic distributions, mass distributions, and charge distributions show good agreement with experimental measurements. The improved DNS-sysu model enables reasonable characterization and description of the QE/grazing collisions, notably resolving long-standing underestimation in the QE channel.
\end{abstract}

\maketitle
$\emph{Introduction.}$ 
The complexity of multinucleon transfer (MNT) reactions originates from the coexistence of multiple overlapping processes, most notably quasi-elastic (QE) scattering, deep-inelastic (DI) scattering, and quasi-fission (QF)\cite{adamian2020extend}. 
Experimental observables such as the TKE–mass distribution in conjunction with the TKEL distribution of the reaction products provide Importance probes of these channels and reveal their intricate interplay \cite{Broda_2006, hirayama2024progress, Corradi_2009, perez2023nuclear,HINDE2021103856,PhysRevC.99.014616,PhysRevC.94.054613, LI2020135697, PhysRevC.109.034608}. Various phenomenological and microscopic approaches have been developed to describe these processes, including the Grazing model \cite{WINTHER1994191, WINTHER1995203}, the dinuclear system (DNS) model \cite{PhysRevC.68.034601, NASIROV2005342, PhysRevC.80.067601, Li2003, PhysRevC.102.054613,Wen_2020}, the Langevin-type transport model \cite{PhysRevC.96.024618, zagrebaev2005unified}, the Coupled Master and  Langevin equations \cite{PhysRevC.109.024617, ZHU2024138423}, the quantum molecular dynamics simulation \cite{Yang:2025xly, ZHAO2021136101, LI2018278}, the time-dependent mean-field theory \cite{PhysRevC.88.014614, PhysRevC.109.024614, PhysRevC.96.024625, PhysRevC.98.064609}, and beyond mean-field approaches \cite{PhysRevC.108.054605, zz3y-22fh}.

Among these, the DNS model stands out for its intuitive picture of two touching nuclei forming a interacting dinuclear configuration, a concept first proposed by Volkov \cite{volkov1978DIC}. The treatment of the internuclear distance as a frozen degree of freedom is a reasonable approximation in MNT reactions, where the configuration is sufficiently long-lived to permit extensive nucleon exchange. Owing to this physical picture, the DNS model has been successfully applied to reproduce a variety of experimental observables in MNT reactions \cite{Li_2022, Zhang:2025pxx, PhysRevC.111.024618, PhysRevC.111.024618, PhysRevC.109.054612}, demonstrating remarkable predictive power. Nevertheless, the applicability of the DNS picture becomes questionable in the regime of grazing collisions/QE \cite{PhysRevC.102.054613, Wen_2017}. In such QE events, only a few nucleons are exchanged, and the interacting nuclei remain relatively far apart with a short-lived contact time. Under these conditions, the assumption of a frozen internuclear distance and a well-formed dinuclear configuration no longer holds, leading to poor performance of the DNS model in describing the QE channel.

The primary objective of this work is to develop such a theoretical model capable of consistently characterizing the QE scattering, DI scattering, and QF processes. We aim to improve the DNS model by introducing a nucleon exchange distance that incorporates relaxation behavior, thereby providing a more accurate and comprehensive description of multinucleon transfer reactions in heavy-ion collisions.

%The article is organized as follows. In Sec. II, the theoretical frameworks of the improved DNS model are presented. 

$\emph{Theoretical framework.}$ The DNS-sysu model is based on the master equation that simulates the evolution of the probability distribution of collective degrees of freedom in non-equilibrium processes. The version of model developed in this work is based on the same foundational framework as that of Zhu $et. al$. \cite{PhysRevC.104.044606} and can be regarded as an extension and further development of the approach. The proposed methodology comprises several key components:
\begin{enumerate}[label=\roman*)]
    \item The multi-dimensional potential energy surface (PES) that characterize the full reaction landscape, including the dynamical deformation $\beta_2$, proton number $Z_{1}$, and neutron number $N_{1}$, providing the driving forces for the evolution of the system.
    \item A three-dimensional master equation that governs the time evolution of probability distributions associated with various nuclear reaction channels, allowing for a quantitative description of the non-equilibrium dynamics in terms of transition rates between different states.
    \item The transfer rates that quantify the likelihood of exchanging nucleons between the interacting nuclei depend not only on the system size and the excitation energy during the reaction \cite{saiko2022multinucleon}, but also on the collision parameters.
    %distance between the two nuclei \cite{zagrebaev2005unified}.
    \item The deflection function method \cite{wolschin1978analysis} provides the model with a reaction time that depends on the deflection parameter.
    %, thereby eliminating the need to introduce additional dynamical parameters or perform complex time-dependent simulations. 
\end{enumerate}

The PES describes the potential energy values that correspond to all possible configurations of a reaction system. By analyzing its geometric features (including minima, saddle points, and gradients), one can gain intuitive understanding of how the system evolves between different states. The PES in the DNS-sysu model can be expressed as:
\begin{flalign}\label{PES}
\begin{aligned}
  U\left(Z_{1}, N_{1}, \beta_{2}, J, R_{\text {cont}}\right)&= \Delta\left(Z_{1}, N_{1}\right)+\Delta\left(Z_{2}, N_{2}\right) \\
  &+V_{\mathrm{cont}}\left(Z_{1}, N_{1}, \beta_{2}, J, R_{\text {cont}}\right)\\
  &+\frac{1}{2} C_{1}\left(\delta \beta_{2}^{1}\right)^{2}
  +\frac{1}{2} C_{2}\left(\delta \beta_{2}^{2}\right)^{2}.
\end{aligned}
\end{flalign}
Here, $\Delta\left(Z_{i}, N_{i}\right)$ ($i$ =1,2) is the mass excess of the $i$th fragment \cite{zhu2018shell}. The last two terms are dynamical deformation energies of the projectile-like fragment (PLF) and target-like fragment (TLF). We restricted the consideration of only one dynamic deformation parameter $\beta_{2}$ instead of two independent deformation $\beta^{1}_{2}$ and $\beta^{2}_{2}$ \cite{Zagrebaev_2007}. The dynamical deformation for each fragment can be calculated by $C_{1}\delta\beta^{1}_{2}=C_{2}\delta\beta^{2}_{2}$ and $\delta\beta^{1}_{2} + \delta\beta^{2}_{2}=2\beta_{2}$. The quadrupole deformation parameters of PLF and TLF are given by $\beta^{1}_{2}=\beta^{\mathrm{p}}_{2}+\delta\beta^{1}_{2}$ and $\beta^{2}_{2}=\beta^{\mathrm{t}}_{2}+\delta\beta^{2}_{2}$, respectively. $\beta^{p}_{2}$ and $\beta^{t}_{2}$ are static quadrupole deformation parameters of the projectile and target, respectively. $C_{1,2}$ are the LDM stiffness parameters of the fragments \cite{MYERS19661}. For the $\beta_{2}$, we take the range of -0.5 to 0.5. The evolution step length is 0.01. The range of \(Z_1\) (\(N_1\)) is from 1 to \(Z_{\mathrm{tot}}\) (\(N_{\mathrm{tot}}\)), with a step size of 1. $R_{\text {cont}}$ is the position where the nucleon transfer process takes place. In the present work, this quantity has been modified to account for its dependence on the collision parameter, exhibiting a relaxation behavior that will be discussed in detail later.
The effective nucleus-nucleus interaction potential consists of the long-range Coulomb repulsive potential, the attractive short-range nuclear potential, and the centrifugal potential:
\begin{flalign}
\begin{aligned}
  V\left(Z_{1}, N_{1}, \beta_{2}, J, r \right) &=V_{\mathrm{N}}\left(Z_{1}, N_{1}, \beta_{2}, r\right) \\
  &+V_{\mathrm{C}}\left(Z_{1}, N_{1}, \beta_{2}, r\right)+\frac{J(J+1) \hbar^{2}}{2 \zeta_{\text {rel }}}.
\end{aligned}
\end{flalign}
Here, $\zeta_{\text {rel}}$ is the moment of inertia for the relative motion of the DNS. More detailed description of Coulomb potential $V_{\mathrm{C}}$ and nuclear potential $V_{\mathrm{N}}$ can be seen in Refs. \cite{PhysRevLett.31.766, ADAMIAN1996, li2005deformation}.

\begin{figure}[htpb]
    \centering
        \includegraphics[width=8.5cm]{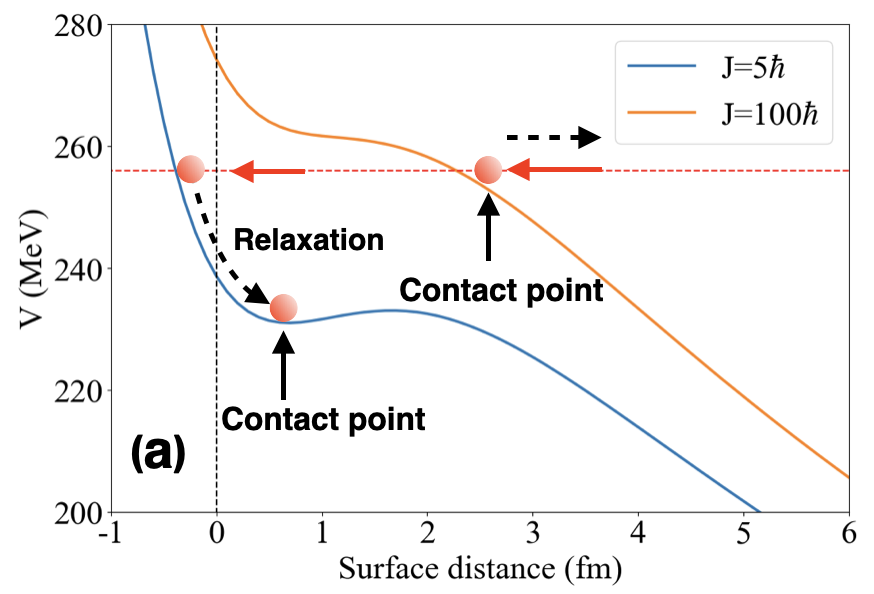}
        \includegraphics[width=8.5cm]{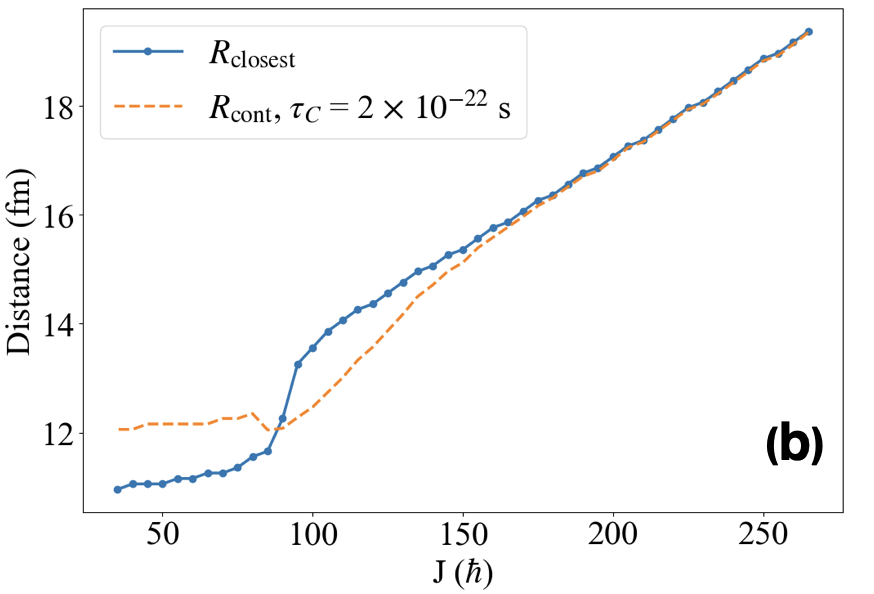}
    \caption{Top panel: Contact points for the $^{58}$Ni + $^{208}$Pb system at $J=5$ $\hbar$ and 100 $\hbar$. The horizontal dashed line represents the incident energy, while the vertical dashed line indicates the sum of the radii of the two nuclei. Bottom panel: The contact position as a function of the entrace angular momentum.} \label{Fig_Vr}
\end{figure}

The MNT reactions involve multiple reaction channels, including QE scattering, DI scattering, and QF processes, which represent distinct dynamical evolution pathways. 
%From the perspective of internuclear distance evolution, 
The intensity of nucleon exchange between the nuclei is expected to correlate with the distance between colliding partners. For the QE scattering/grazing collisions, the nuclear potential exerts a relatively weak attraction on the projectile, causing its trajectory to closely follow that of Rutherford scattering. As a result, the two nuclei remain at relatively large distances throughout the interaction and the relaxation time is rather short. And the distance of closest approach $R_{\mathrm{closest}}$ can be obtained by equilibrating the incident energy and the interaction potential $V(Z_{\mathrm{pro}}, N_{\mathrm{pro}}, \beta_{2}=0, J,r)$. As the reaction progresses toward more violent collisions, usually corresponding to smaller impact parameters, the internuclear separation decreases significantly. Due to the Pauli exclusion principle, the internuclear distance cannot become too small. Instead, nucleon exchange typically occurs with a long relaxation time when the system resides at a relatively flat region of the potential energy surface, such as near the bottom of the potential pocket ($R_{\mathrm{bottom}}$). For systems without a potential pocket, this value $R_{\mathrm{bottom}}$ is taken at a surface separation of approximately 0.7 fm. Therefore, in this work, the contact distance between the nuclei is represented phenomenologically by introducing a relaxation behavior:
\begin{flalign}
\begin{aligned}
R_{\text {cont}} = R_{\mathrm{closest}}f(t) + R_{\mathrm{bottom}}[1-f(t)].
\end{aligned}
\end{flalign}
Here $t$ is the interaction time at each angular momentum, obtained through the deflection function \cite{wolschin1978analysis}. $f(t)=\mathrm{ \exp}{(-t/\tau_{C})}$ is a smoothing function with characteristic time $\tau_{\mathrm{C}}=2 \times 10^{-22} \mathrm{s}$. 
The relaxation time can be determined from the analysis of experimental data. 

In the case of QE scattering, such as when the relative angular momentum is around 100 $\hbar$ as shown in the Fig. \ref{Fig_Vr}(a), the extremely short interaction time (\(t \leq 10^{-22}\,\mathrm{s}\)) prevents any significant relaxation. After reaching the point of closest approach, the nuclei are immediately scattered apart. However, as the collision becomes more violent, strong friction forces and large nuclear viscosity in the entrance channel drive the system toward relaxation, gradually settling into the flatter region of the adiabatic potential where nucleon exchange occurs. These two distinct dynamical regimes are schematically illustrated in Fig. \ref{Fig_Vr} (a). 
In Fig. \ref{Fig_Vr}(b), the blue line represents the closest approach distance between the nuclei as a function of angular momentum, while the red dashed line includes the effect of relaxation behavior. It can be observed that in the region of large angular momentum (the QE scattering case), the difference between the two curves is negligible. However, at low angular momentum, the relaxed contact distance tends to stabilize around a specific value.

The master equation (ME) describes the evolution of the probability distribution of states of a continuous-time Markov process \cite{Haag2017ModellingWT2}. The general form of the ME can be written as:
\begin{flalign}
\begin{aligned}
\frac{\partial P(\mathbf{S}, t)}{\partial t} = \sum_{\mathbf{S}^{\prime} \neq \mathbf{S} } & W(\mathbf{S}^{\prime},\mathbf{S})P(\mathbf{S}^{\prime}, t)  - W(\mathbf{S},\mathbf{S}^{\prime})P(\mathbf{S},t).
\end{aligned}
\end{flalign}
$P(\mathbf{S}, t)$ denotes the occupation probability of state $\mathbf{S}$. The quantitiy $W$ is transition probability per time unit or transition rate in the following sense: $W(\mathbf{S}^{\prime},\mathbf{S})P(\mathbf{S}^{\prime}, t)$ is the probability transferred from state $\mathbf{S}^{\prime}$ to the state $\mathbf{S}$ per unit of time.

In the DNS-sysu model, the number of proton $Z_{1}$, the number of neutron $N_{1}$, and the dynamical deformation ${\beta_{2}}$ degree of freedom for describing the macroscopic observables of PLF. Hence, $\mathbf{S}=(Z_{1}, N_{1}, \beta_{2})$. In this model, each time step permits only a single transition to an adjacent macroscopic state, corresponding to a one-unit change in either \(Z_1\) or \(N_1\), with the spatial scale of the macroscopic states precisely matching the potential energy surface grid defined in the previous section. Therefore, the joint probability of fragments $P$ evolving with different macroscopic degrees of freedom can be solved numerically as follows:
\begin{widetext}
\begin{flalign}
\begin{aligned}
\frac{\partial P(\beta_2,Z_1,N_1,t)}{\partial t} &=  W(\beta_2-\Delta \beta_2,\beta_2)P(\beta_2-\Delta \beta_2,Z_1,N_1,t) + W(\beta_2+\Delta \beta_2,\beta_2)P(\beta_2+\Delta \beta_2,Z_1,N_1,t) 
\\& + W(Z_1-\Delta Z,Z_1)P(\beta_2,Z_1-\Delta Z,N_1,t) + W(Z_1+\Delta Z,Z_1)P(\beta_2,Z_1+\Delta Z,N_1,t) 
\\& + W(N_1-\Delta N,N_1)P(\beta_2,Z_1,N_1-\Delta N,t) + W(N_1+\Delta N,N_1)P(\beta_2,Z_1,N_1+\Delta N,t) 
\\& - [W(\beta_2,\beta_2+\Delta \beta_2) + W(\beta_2,\beta_2-\Delta \beta_2)]P(\beta_2,Z_1,N_1,t) 
\\& - [W(Z_1,Z_1+\Delta Z) + W(Z_1,Z_1-\Delta Z)]P(\beta_2,Z_1,N_1,t) 
\\& - [W(N_1,N_1+\Delta N) + W(N_1,N_1-\Delta N)]P(\beta_2,Z_1,N_1,t).
\end{aligned}
\end{flalign}
\end{widetext}

Here, $W(Z_1-\Delta Z,Z_1)$ represents the macroscopic transition rate from the channel ($Z_1-\Delta Z$, $N_1$, $\beta_2$) to ($Z_1$, $N_1$, $\beta_2$), and the other macroscopic transition rates are similar. A single transition is restricted to interactions between adjacent macroscopic states; that is, transitions occur with $\Delta Z = 1$, $\Delta N = 1$, and $\Delta \beta_2 = 0.01$.

The transfer rate $W$ serves as a critical bridge between microscopic nucleon exchange mechanisms and macroscopic observables. For nuclei in contact the macroscopic transition rate can be treated by \cite{moretto1975theoretical, zagrebaev2005unified}:
\begin{flalign}
\begin{aligned}
 W(\mathbf{S}^{\prime},\mathbf{S}) \propto \lambda_{0} \exp(\frac{U(\mathbf{S}^{\prime})- U(\mathbf{S})}{2T}).
\end{aligned}
\end{flalign}

Although, $\lambda_{0}$ ($\approx$ $10^{22} \mathrm{s}^{-1}$) serves as the nucleon transfer rate in the model, its derivation from first principles, particularly its coupling to dissipation dynamics and potential energy gradients, has yet to be rigorously established. As usual, it can be qualitatively assumed that it is related to the size of the system and the temperature \cite{moretto1975theoretical, zagrebaev2015cross}. In this study, we adopt the nucleon transfer rate identical to that ($5 \times \mathrm{A^{2}_{tot}} \times \mathrm{(T/MeV)} \times 10^{16}$) reported in Ref. \cite{,saiko2022multinucleon}, ensuring consistency with established theoretical benchmarks for comparative analysis.

The term $\exp(\frac{U(\mathbf{S}^{\prime})- U(\mathbf{S})}{2T})$ governs the thermally driven transitions between macroscopic states $\mathbf{S}^{\prime}$ and $\mathbf{S}$ in the master equation. Here, $U$($\mathbf{S}$) denotes the potential energy of the collective coordinates ($e.g.$, fragment($Z_1$, $N_1$), dynamical deformation $\beta_2$) , while $T$ represents the local nuclear temperature $\sqrt{E^*/\alpha}$, $E^*$ is the excitation energy, $\alpha$ = $A_{\mathrm{tot}}$/12 $\mathrm{MeV^{-1}}$ is the level density parameter.

For separated nuclei, nucleon exchange/transfer could still take place depending on the extension of the tails of the single particle wavefunctions. %Nucleons can be still transferred between nuclei that are somewhat far apart. 
This happens mainly through temporary connections (necks) between the nuclei, which is important for grazing collisions. 
%We need to include this effect when we calculate how nucleons move between nuclei. 
Therefore, the transition rate $W$ can be further written as the following expression \cite{zagrebaev2003sub, von1987quasi}:
\begin{flalign}
\begin{aligned}
 W(\mathbf{S}^{\prime},\mathbf{S}) = \lambda_{0} \exp(\frac{U(\mathbf{S}^{\prime})- U(\mathbf{S})}{2T})P_{\mathrm{tr}}(R_{\mathrm{cont}}, \mathbf{S} \rightarrow \mathbf{S} \pm \Delta\mathbf{S}).
\end{aligned}
\end{flalign}

\begin{figure}[htpb]
    \centering
        \includegraphics[width=8.5cm]{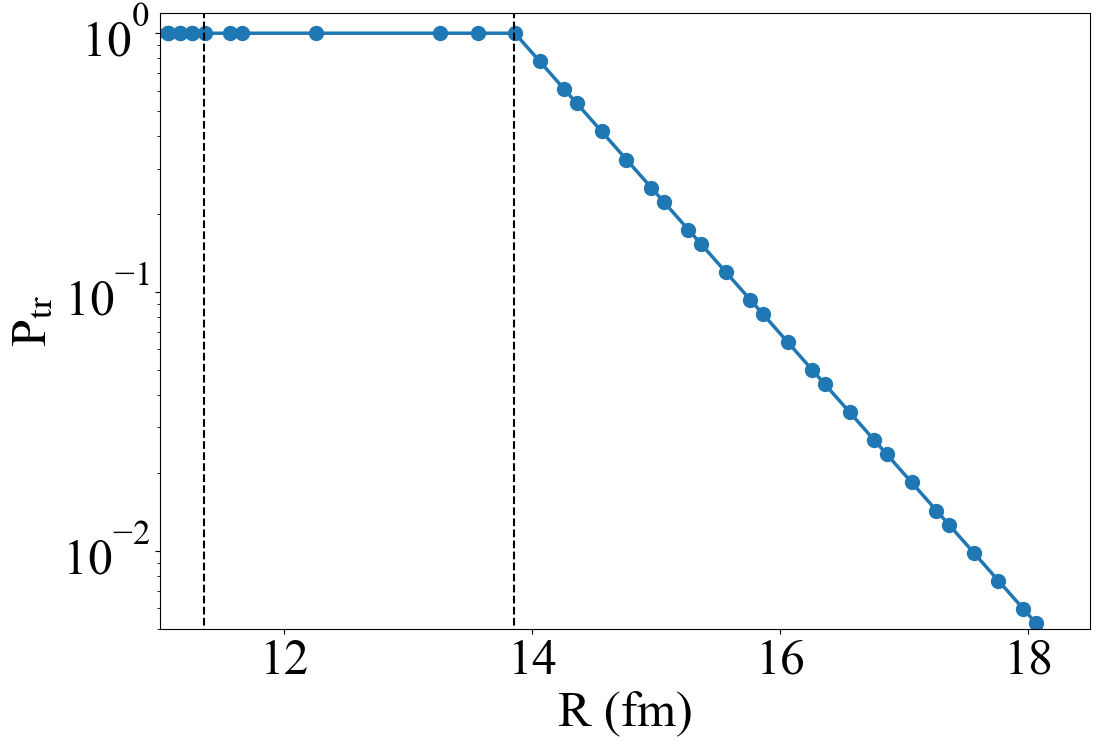}
    \caption{The nucleon transfer probability as a function of the center distance in the reaction $^{58}$Ni + $^{208}$Pb at $E_\mathrm{c.m.}$ = 256 MeV.} \label{Fig_Ptr}
\end{figure}

Here, $P_{\mathrm{tr}}(R_{\mathrm{cont}}, \mathbf{S} \rightarrow \mathbf{S} \pm \Delta\mathbf{S})$ is the probability of macroscopic degrees of freedom $\mathbf{S}$ depends on the distance between the nuclear surfaces. The semiclassical approximation $\exp(-2k[R-R_{\mathrm{tr}}])$ is used for calculating $P_{\mathrm{tr}}$. $R_{\mathrm{tr}}= R_{\mathrm{pro}} + R_{\mathrm{tar}} + 2.5$ fm. For the neutron, $k = \sqrt{M(-\epsilon_{F})/2\hbar^{2}} + \sqrt{M(-\epsilon^{\prime}_{F})/2\hbar^{2}}$. For the proton, $k = \sqrt{M(-\epsilon_{F}+Z_{T}e^{2}/R_{T})/2\hbar^{2}} + \sqrt{M(-\epsilon^{\prime}_{F}+Z_{P}e^{2}/R_{P})/2\hbar^{2}}$. $M$ means the nucleon mass and $\epsilon_{F}$ ($\epsilon^{\prime}_{F}$) denotes the nucleon separation energy of projectile (target). Figure \ref{Fig_Ptr} shows the nucleon transfer probability as a function of the internuclear separation \(R\). Each data point corresponds to a simulation performed at the indicated collision angular momentum. The left vertical dashed line denotes the sum of the radii of the two nuclei, while the right one marks a surface separation of \(2.5\ \mathrm{fm}\). This probability goes exponentially to zero at $R\rightarrow \infty $ and it is equal to unity when $R$ is smaller than $R_{\mathrm{tr}}$. 

%The key parameters of the model are summarized in Table I. These parameters govern the time evolution of the relaxation processes and the strength of nucleon transfer. Their values were determined through a systematic experiment comparison. This set of parameters enables a reasonable reproduction of the available experimental data for MNT reactions, ranging from $^{40}$Ca + $^{208}$Pb up to $^{208}$Pb + $^{208}$Pb.

\begin{figure}[htpb]
    \centering
        \includegraphics[width=8.0cm]{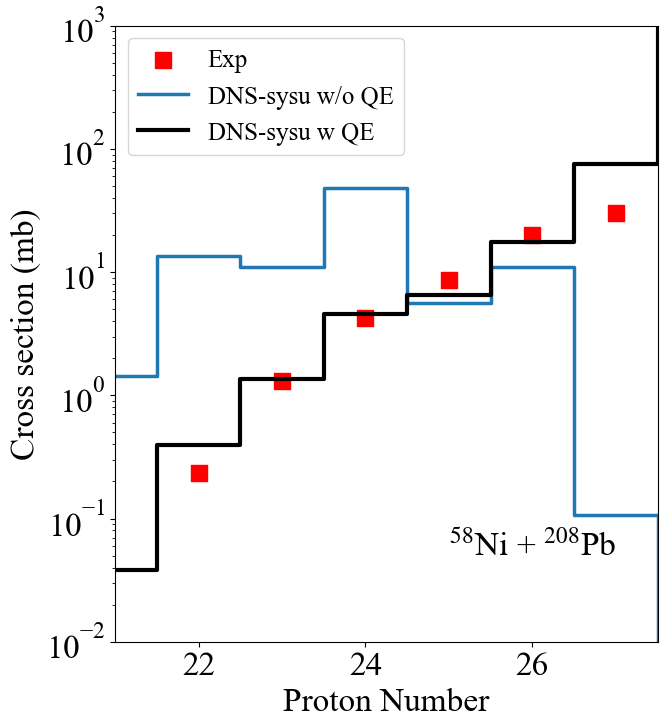}
    \caption{Total cross sections for pure proton stripping channels in the reaction $^{58}$Ni + $^{208}$Pb at $E_\mathrm{c.m.}$ = 256 MeV. The blue line represents the DNS-sysu result without the QE channel. The black solid line represents the result with the QE channel.} \label{figure3}
\end{figure}

$\emph{Results and discussions.}$
Figure \ref{figure3} shows the comparison of calculated charge distributions with the experimental data \cite{PhysRevC.66.024606} for the MNT reaction $^{58}$Ni + $^{208}$Pb at $E_\mathrm{c.m.}$ = 256 MeV. The code GEMINI++ \cite{PhysRevC.82.014610} is used to treat the de-excitation process of primary fragments. 

Fig. \ref{figure3} illustrates the significant improvement achieved by the DNS-sysu model with the QE channel in reproducing the experimental transfer cross sections, particularly for few-nucleon transfer processes. Note that experimental detection conditions are applied as constraints in the calculations. Specifically, we implement an angular momentum cutoff to select computational results that fall within the experimentally observable range. The DNS-sysu result without the QE channel clearly demonstrates its failure to describe and significant underestimation of the experimental data for one to two-proton transfer reactions. As discussed previously, this deficiency stems from the model's inherent limitation in accounting for long-range nucleon transfer probabilities. In contrast, the improved model shows excellent agreement with the few-nucleon transfer cross sections, successfully reproducing both the absolute values and the slope of the cross-section decrease as a function of transferred nucleon number.

\begin{figure*}[htpb]
    \centering
        \includegraphics[width=16cm]{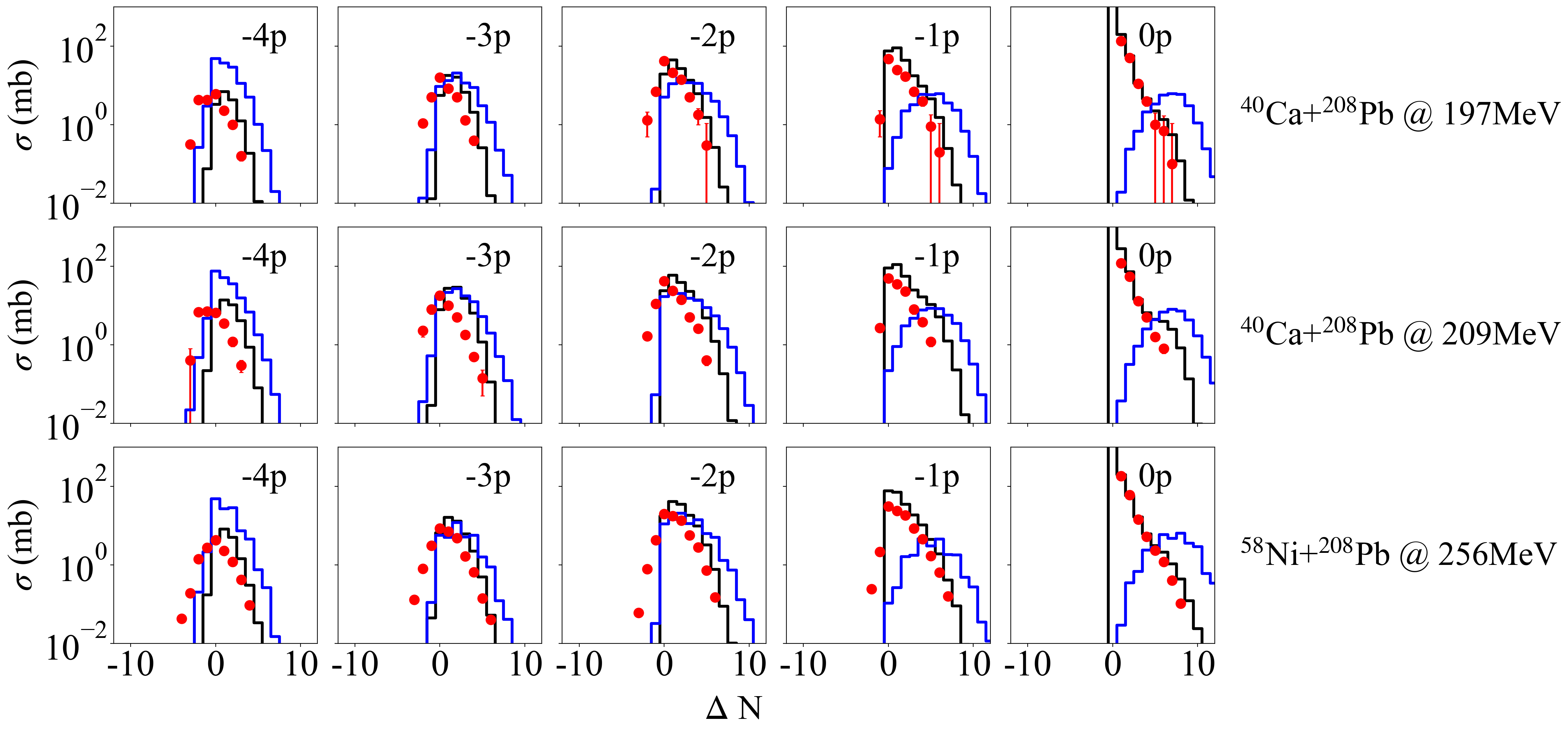}
    \caption{Experimental cross sections for $^{40}$Ca + $^{208}$Pb (the first and second row) and $^{58}$Ni+ $^{208}$Pb (the third row), at energies $E_\mathrm{c.m.}$ = 197, 209, and 256 MeV, respectively. The black and blue solid lines denote the calculated results with the improved and the origin DNS model, respectively.} \label{figure4}
\end{figure*}
In order to further test the improved DNS-sysu model, in Fig. \ref{figure4}, we present a systematic comparison of isotopic distributions between experimental data and calculations for four representative reaction systems: beams $^{40}$Ca and $^{58}$Ni incident on a $^{208}$Pb target at center-of-mass energies of E$_{\mathrm{c.m.}}$ = 197, 209, and 256 MeV \cite{PhysRevC.94.064616, PhysRevC.66.024606, PhysRevC.71.044610}, respectively. 
%The $^{208}$Pb target was specifically selected for this comparative study due to its doubly magic nature, which provides enhanced theoretical reliability, and its minimal susceptibility to competing mechanisms such as fission processes that could complicate the interpretation of MNT observables.
The calculations show excellent agreement with experimental data for the (0p) channel, accurately reproducing neutron pick-up cross sections. However, minor deviations emerge in the (-1p) channel, where the model slightly overestimates neutron pick-up cross sections while modestly underestimating neutron stripping values. These discrepancies become progressively more pronounced in the (-2p), (-3p), and (-4p) channels, demonstrating a systematic trend where the difference between theory and experiment grows with increasing number of protons removed from the projectile.

Moreover, the most significant difference between the original model and the improved one lies in the results for (0p) and (-1p) transfer. It is evident that the improved DNS-sysu model with consideration of angular momentum dependent transition probability strongly improve the description of QE/grazing collisions.
%fails to reproduce the exponential decrease of the cross section with increasing neutron transfer. This behavior essentially results from changes in the transfer probability due to varying distances between the nuclei, which goes beyond the assumptions made in the original model.

\begin{figure*}[htpb]
    \centering
        \includegraphics[width=15cm]{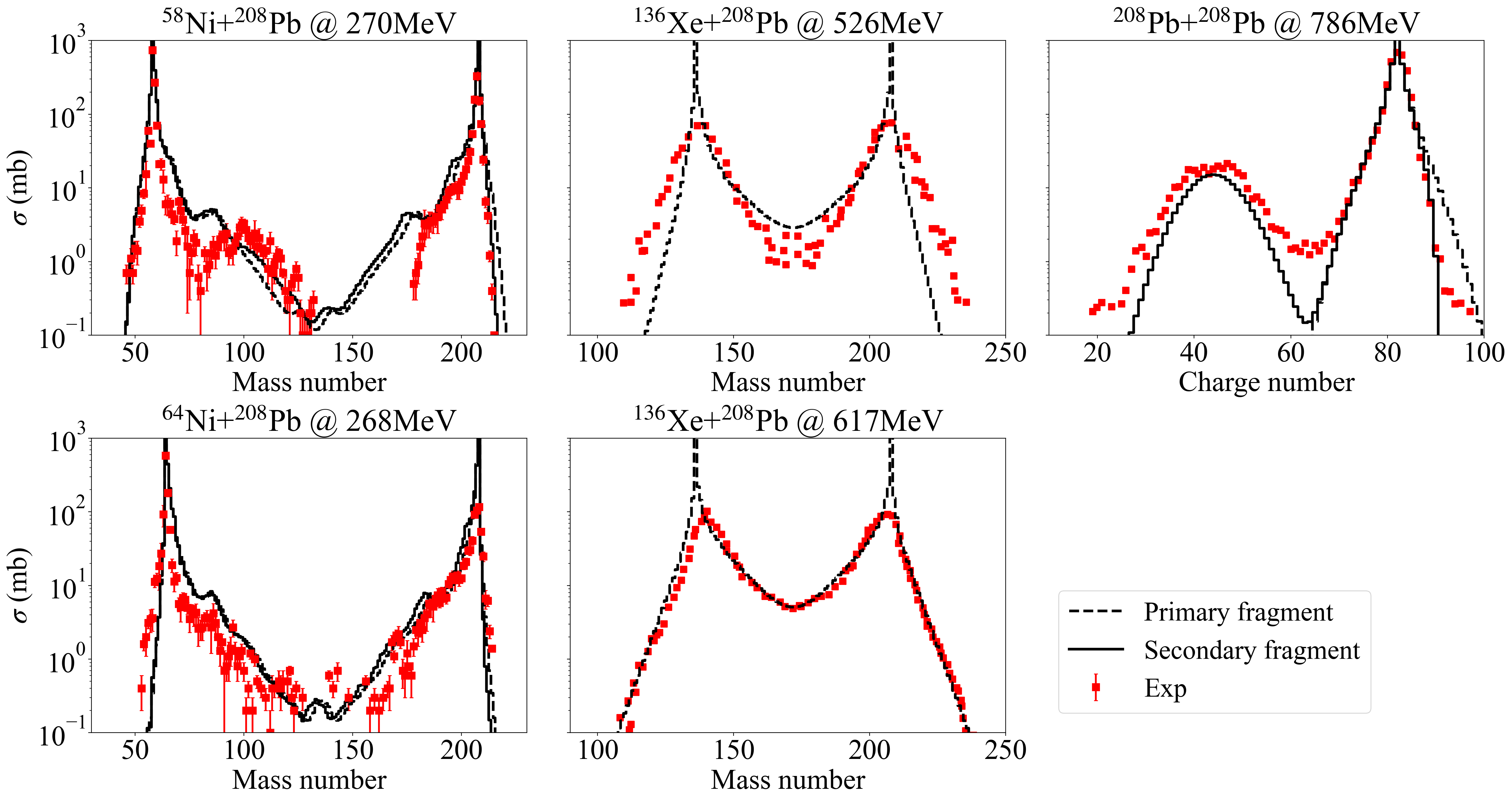}
    \caption{Mass or charge distributions for reaction products. The measured cross sections are shown by red squares with error bars, which were taken from Refs \cite{KROLAS2010170, KROLAS2003289, PhysRevC.86.044611, PhysRevC.86.044611, TANABE1980194}. Black dotted line represents the results of the improved DNS-sysu calculations, while the black solid line represents the results of the improved DNS-sysu + GEMINI++ calculations.} \label{figure5}
\end{figure*}

In addition to comparing isotopic cross sections that directly reflect nucleon transfer patterns, we also systematically examine the distributions of reaction products that provide more intuitive characterization of collective behavior in MNT processes where massive nucleon redistribution occurs.
The comparison of the available experimental data on mass or charge distributions formed in various MNT reactions is shown in Fig. \ref{figure5} for the following reactions: $^{58}$Ni + $^{208}$Pb at $E_\mathrm{c.m.}$ = 270 MeV \cite{KROLAS2010170}, $^{64}$Ni + $^{208}$Pb at $E_\mathrm{c.m.}$= 268 MeV \cite{KROLAS2003289}, $^{136}$Xe + $^{208}$Pb at $E_\mathrm{c.m.}$ = 526 MeV \cite{PhysRevC.86.044611}, $^{136}$Xe + $^{208}$Pb at $E_\mathrm{c.m.}$ = 617 MeV \cite{PhysRevC.86.044611}, and $^{208}$Pb + $^{208}$Pb at $E_\mathrm{c.m.}$ = 786 MeV \cite{TANABE1980194}. 
One can see from each panel of Fig. \ref{figure5} that the calculated distributions of the production cross sections are reasonably consistent with experimental results.

%The results of the secondary production cross sections (black solid lines) are compared with available experimental data (red squares). The experimental data for $^{136}\mathrm{Xe} + ^{208}\mathrm{Pb}$ correspond to the cross sections of the primary fragment \cite{PhysRevC.86.044611}, while the data for $^{208}\mathrm{Pb} + ^{208}\mathrm{Pb}$ have been rescaled in magnitude, as the values reported in the literature are given in arbitrary units \cite{TANABE1980194}. 

$\emph{Summary.}$ In this Letter, we have developed the DNS-sysu model aimed at providing a comprehensive and unified description of QE, DI, and QF channels in MNT reactions. The improved DNS-sysu model presented here incorporates a multidimensional master-equation-based diffusion approach, which allows for a realistic simulation of the stochastic nucleon exchange process during the interaction of two heavy nuclei. The validity and predictive power of the model were benchmarked against a broad set of experimental data, covering reactions such as $^{40}$Ca, $^{58, 64}$Ni, $^{136}$Xe, and $^{208}$Pb + $^{208}$Pb. The calculated isotopic, mass, and charge distributions of the reaction products exhibit strong consistency with the experimental measurements. 
One of the key improvements of the DNS model in this work is the inclusion of partial impact-parameter-dependent transition probabilities, which enhances the description of both grazing collisions and DIC. %By extending the model's ability to treat QE scattering, DIC, and QF within a single framework, we provide a more realistic and flexible tool for exploring nuclear reaction dynamics. 
This unified approach is particularly valuable for identifying optimal reaction systems and beam energies to access unexplored regions of the nuclear chart.

%From a theoretical perspective, the DNS model remains a semi-microscopic approach, striking a balance between computational tractability and physical realism. While it does not resolve all microscopic degrees of freedom, such as single-particle motion or time-dependent shape evolution, it captures the essential collective behavior of nucleon exchange and dissipation in a statistically meaningful way. This makes it especially suited for large-scale surveys of nuclear reactions where full microscopic models ($e.g.$, time-dependent Hartree-Fock or quantum molecular dynamics) are computationally prohibitive.

Nevertheless, there remain open challenges and opportunities for future development. One possible direction is the integration of microscopic inputs, such as single-particle energy levels and nucleon density distributions from constrained density functional theory, to provide more accurate potential energy surfaces and transition rates. 

$\emph{Acknowledgments.}$ The authors would like to thank Pei-Wei Wen and Gen Zhang for helpful discussion and suggestions. This work was supported by the National Natural Science Foundation of China under Grants No. 12075327 and 12475136.

%\tableofcontents
\bibliography{ref}

\end{document}